\documentclass[preprint,showpacs,preprintnumbers,amsmath,amssymb]{revtex4}
\usepackage{graphicx}
\usepackage{dcolumn}
\usepackage{bm}

\begin{document}

\title{Short DNA persistence length in a mesoscopic helical model}

\author{Marco Zoli}

\affiliation{School of Science and Technology \\  University of Camerino, I-62032 Camerino, Italy \\ marco.zoli@unicam.it}

\begin{abstract}
The flexibility of short DNA chains is investigated via computation of the average correlation function between dimers which defines the persistence length.  Path integration techniques have been applied to confine the phase space available to base pair fluctuations and derive the partition function. The apparent persistence lengths of a set of short chains have been computed as a function of the twist conformation both in the over-twisted and the untwisted regimes, whereby the equilibrium twist is selected by free energy minimization. The obtained values are significantly lower than those generally attributed to kilo-base long DNA. This points to an intrinsic helix flexibility at short length scales, arising from large fluctuational effects and local bending, in line with recent experimental indications. 
The interplay between helical untwisting and persistence length has been discussed for a heterogeneous fragment by weighing the effects of the sequence specificities through the non-linear stacking potential.
\end{abstract}


\maketitle

The DNA double helical structure is stable enough to preserve genetic information encoded in the Watson-Crick paired bases and also loose enough to allow for those transient base pair (bp) openings \cite{zocchi03} which make the code accessible to enzymes during the processes of replication, transcription and repair. 
At physiological temperatures DNA molecules fluctuate between a variety of random coil conformations in which even distant segments along the helical axis can be brought in close proximity \cite{biton18}. This points to an inherent flexibility of the DNA chain which has been widely probed over the last twenty five years \cite{busta92}. 
While these experiments demonstrate that stretch and twist elasticity are intertwined \cite{bohr11}, they also call for models in which bp fluctuations and stacking interactions are considered as dependent on the specific twist conformation of the molecule.  Modeling of the helix and its conformational states can be carried out at different levels of resolution ranging from all-atom simulations to continuous worm-like chain (WLC) models which simply treat DNA as a homogeneous and inextensible rod \cite{oroz16}, not accounting for the interplay between twist and bp fluctuations.  This may explain the shortcomings of the WLC model emerged in the analysis of the cyclization properties \cite{vafa,io16b} at those short length scales in which the details of the bp interactions matter. 
In this regard, mechanical models such as the Dauxois-Peyrard-Bishop (DPB) model \cite{pey93b} provide a convenient description of the dsDNA in which  the complementary strands are represented by two parallel chains of beads coupled via a intra-chain anharmonic potential.  However, the  standard mesoscopic modeling has the general drawback that the twist and bending conformational degrees of freedom are frozen (or absent) when the base pairs (bps) vibrate. 
To overcome this limitation we have proposed a model which includes the angular variables in the intra-chain stacking interactions and developed a method to determine the equilibrium twist conformation of short homogeneous oligomers. 
The efficacy of the method has been recently tested by evaluating the DNA elastic response in the presence of a stretching perturbation \cite{io18} while previous studies had examined the helix unwinding and formation of denaturation bubbles in circular DNA as a function both of temperature and of the circle size \cite{io13,io14a}. 

Here we focus on a key indicator of the polymer flexibility, namely its persistence length ($l_p$), which measures the orientational correlation between distant segments of the chain and, for a discrete model, it can be calculated as a sum over the average scalar products of the bond vectors associated to those segments \cite{soder97}. While this microscopic approach proves useful to deal with the end effects associated to short oligomers, such correlation distance depends on the chain length and contains electrostatic contributions arising from the fact that distant monomers along the molecule stack may be brought close to each other because of bending fluctuations. Then, our microscopic correlation distance provides a measure of the \textit{apparent} $l_p$  conceptually distinct from the \textit{intrinsic} $l_p$ which instead defines a local property of the polymer, independent of the chain length. On the other hand, the \textit{apparent} $l_p$ is also the one which can be compared with the experiments, as it is generally extracted from measurements of global properties of the helical molecule e.g., the end-to-end distance obtained by fluorescence
resonance energy transfer (FRET) \cite{archer08}. It is also noticed that the methods used to extract $l_p$'s data rely on global equations which, strictly speaking, have been derived in the framework of  WLC continuous models for kilo-base long polymers and whose application to short length scales may be questionable \cite{maiti15}. For these reasons, we pursue here a research line alternative to the WLC approach and present the theoretical background to calculate the $l_p$'s at short length scales on the base of a mesoscopic Hamiltonian containing the forces which stabilize the molecule.

\section{Model }

\begin{figure}
\includegraphics[height=6.0cm,width=8.0cm,angle=-90]{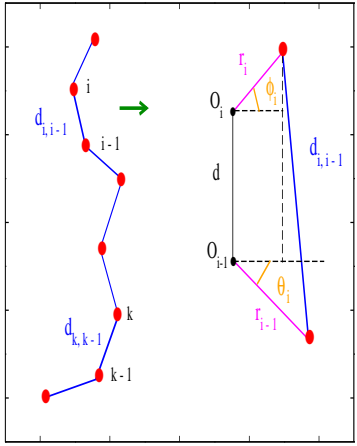}
\caption{\label{fig:1}(Color online)  
Open end chain with $N$ base pairs (red dots) represented by the relative distances $r_i$'s between the two pair mates on complementary strands.  The $r_i$'s are measured with respect to the mid-axis of the helix and represent the radial fluctuations with respect to the bare helix diameter. 
$\theta_i$ and $\phi_{i}$ are the twisting and bending angles respectively formed by adjacent $r_{i}$ and $r_{i-1}$ along the molecule stack. The bond vector $d_{i,\,i-1}$ denotes the  separation between adjacent base pairs. 
}
\end{figure}

We consider a model for the helix with $N$ bps as depicted in Fig.~\ref{fig:1} in which the stretching vibrations between the mates of the $i-th$ bp are defined by, $\, r_{i}=\,r_{i}^{(2)} - r_{i}^{(1)}$, where $r_{i}^{(1,2)}$ are the positions of the complementary mates, respectively given by: $r_{i}^{(1)}=\, -R_0/2 + x_{i}^{(1)}$ and $r_{i}^{(2)}=\, R_0/2 + x_{i}^{(2)}$.  $x_{i}^{(1,2)}$ represent the fluctuations of the two bases in the pair and $R_0$ is the inter-strands separation in the absence of radial fluctuations i.e., the bare helix diameter. Suppressing the radial fluctuations, all $r_{i}=\,R_{0}$,  the red dots in Fig.~\ref{fig:1} would map onto the the $O_i$'s lying along the molecule mid-axis at a constant rise distance $d$. The latter is the bond length in the freely jointed chain model. Hereafter, the bare helix parameters are set to the average values usually assumed for kilo-base long DNA, i.e., $R_0 = \,20 $\AA {} and $d = \, 3.4$\AA {}. 

As shown in the right panel of Fig.~\ref{fig:1},  adjacent bps along the chain are twisted by 
$\theta_i$ and bent by $\phi_{i}$. Thus, the sequence specific distance $\overline{d_{i,i-1}}$ between  $r_{i}$ and $r_{i-1}$, which can be straightforwardly obtained by geometrical arguments, depends both on the radial and angular fluctuations.
Hence, the computation of average distances $< \overline{d_{i,i-1}} >$ and average scalar products $< \overline{d_{i,i-1}} \cdot \overline{d_{k,k-1}}>$ involves  integrations over the whole ensemble of radial and angular fluctuations consistent with the model potential.

The helical molecule is described by a potential given by the sum of \textit{i)} a one particle term $V_{1}[r_i]$ which models the hydrogen bonds between complementary pair mates and \textit{ii)} a two particles term $V_{2}[ r_i, r_{i-1}, \phi_i, \theta_i]$ which models the stacking forces essentially due to the overlap of $\pi$ electrons on adjacent bases along the stack. It is the stacking which depends on the twisting and bending variables. Analytically the potential reads:

\begin{eqnarray}
& &V_{1}[r_i]=\, V_{M}[r_i] + V_{Sol}[r_i] \, , \nonumber
\\
& &V_{M}[r_i]=\, D_i \bigl[\exp(-b_i (|r_i| - R_0)) - 1 \bigr]^2  \, , \nonumber
\\
& &V_{Sol}[r_i]=\, - D_i f_s \bigl(\tanh((|r_i| - R_0)/ l_s) - 1 \bigr) \, , \nonumber
\\
& &V_{2}[ r_i, r_{i-1}, \phi_i, \theta_i]=\, K_S \cdot \bigl(1 + G_{i, i-1}\bigr) \cdot \overline{d_{i,i-1}}^2  \, , \nonumber
\\
& &G_{i, i-1}= \, \rho_{i, i-1}\exp\bigl[-\alpha_{i, i-1}(|r_i| + |r_{i-1}| - 2R_0)\bigr]  \, \nonumber
\\. 
\label{eq:01}
\end{eqnarray}

The one particle term is the sum of a Morse potential $V_{M}[r_i]$ and a solvent potential $V_{Sol}[r_i]$.
$V_{M}$ represents an effective pair interaction energy between nucleotides 
measured from the zero level which corresponds to the absence of radial fluctuations, i.e., $|r_i| = R_0$.  While $|r_i|$ may also become smaller than $R_0$, it is important to emphasize that the range of the radial fluctuations is set by the potential parameters i.e., the depth $D_i$ and width $b_i$. In fact, the code includes those bp vibrations such that $V_{M}[r_i] \leq D_i $ which amounts to exclude those displacements such that $|r_i| - R_0 < - \ln 2 / b_i$. The latter are discouraged by the electrostatic repulsion and would yield a negligible contribution to the partition function.
On the other hand, for $|r_i| \gg  R_0$, $V_M$ becomes flat yielding a vanishing force between the bp mates  as expected when
the bases get far apart and pair dissociation sets in.
While the Morse potential is generally suitable to describe the equilibrium properties of DNA, it has to be corrected to account for the DNA dynamics. In fact, when a base flips out of the stack, there is an entropic gain associated to the new available degrees of freedom and the open bases may also form hydrogen bonds with the solvent. Thus, in order to re-close, the base has to overcome an entropic barrier, not described by $V_M$. The latter is introduced in our model by the term $V_{Sol}$ which enhances by the factor $D_i f_s$ the energy threshold for bp breaking above the Morse dissociation energy and provides the barrier, whose width is determined by $l_s$, which accounts for those effects of strand recombination occurring in solution.

The two particles term $V_{2}[ r_i, r_{i-1}, \phi_i, \theta_i]$ contains the elastic constant $K_S$ and the non-linear parameters $\rho_{i, i-1}, \, \alpha_{i, i-1}$ (discussed below).
$V_{2}$ is an extension of the non-linear stacking potential first introduced in the DPB ladder model to account for the first order-like denaturation transition associated to the bp opening and strand separation. In ladder models however,  whenever two adjacent bases slide far away from each other so that the overlap between their $\pi$ electrons is lost, the stacking energy becomes infinitely large. Instead, our potential realistically accounts for the finiteness of the stacking interaction due to the stiffness of the sugar-phosphate backbone. This is done through the angular variables between adjacent bases, in particular through the twist angle which yields a restoring force in the stacking also in the presence of large amplitude fluctuations \cite{io12}.

\section{ Method}

The model potential in Eq.~(\ref{eq:01}) is treated by the finite temperature path integral techniques widely described in some previous works \cite{io11,io14b}.
The idea underlying our method is that of mapping the bp displacements onto the time axis, $r_i \rightarrow |r_i(\tau)|$,  so that the separation between the pair mates is a time dependent trajectory varying in the range $\, \tau_b - \tau_a \,$ whose upper limit is  $\, \beta=\,(k_{B}T)^{-1}$, $T$ being the temperature and $k_B$ the Boltzmann constant.
As the partition function is an integral over closed trajectories, $(\,r_i(0)=\, r_i(\beta) \,)$  running along the $\tau$-axis, the bp trajectories can be expanded in Fourier series whose coefficients generate the large ensemble of radial bp fluctuations contributing with their statistical weight to the averages of the  helical parameters. 

Technically, integrating over the Fourier coefficients, an increasing number of trajectories is added to the partition function $Z_N$ until the latter converges.  As the model also depends on bending and twisting degrees of freedom, the numerical convergence must be obtained also integrating over the angular variables with their specific cutoffs.
This ultimately corresponds to the state of thermodynamic equilibrium which is achieved by summing over about $10^8$ configurations for each dimer in the chain.
The method has essentially the following advantages: \textit{i)} it introduces the $T$ dependence in the formalism;  \textit{ii)} it  sets the cutoffs on the bp amplitudes by defining an integration measure which normalizes the kinetic action (see below). Accordingly  the phase space available to the $r_{i}$'s is consistently confined without operating arbitrary truncations in order to remove the divergence of the partition function as found in Hamiltonian investigations of DNA thermal denaturation \cite{zhang97,munoz10}; 
\textit{iii)} it directly relates the macroscopic helix parameters, e.g. average diameter and rise distance, to the fluctuational effects treated at the level of the base pair.
Then, the partition function $Z_N$ for the chain of $N$ bps of reduced mass $\mu$, is:

\begin{eqnarray}
& &Z_N=\, \oint Dr_{1} \exp \bigl[- A_a[r_1] \bigr]   \prod_{i=2}^{N}  \int_{- \phi_{M} }^{\phi_{M} } d \phi_i \int_{- \theta_{M} }^{\theta _{M} } d \theta_{i} \cdot \,
\nonumber
\\
& &\oint Dr_{i}  \exp \bigl[- A_b [r_i, r_{i-1}, \phi_i, \theta_i] \bigr] \, , \nonumber
\\
& &A_a[r_1]= \,  \int_{0}^{\beta} d\tau \biggl( \frac{\mu}{2} \dot{r}_1(\tau)^2 + V_{1}(\tau) \biggr)\, , \nonumber
\\
& &A_b[r_i, r_{i-1}, \phi_i, \theta_i]= \,  \int_{0}^{\beta} d\tau \biggl( \frac{\mu}{2} \dot{r}_i(\tau)^2 + V_{1}(\tau) + V_{2}(\tau) \biggr) \, , \nonumber
\\
& &V_{1}(\tau)\equiv \,V_{1}[r_i(\tau)] \, ; \, V_{2}(\tau)\equiv \,V_{2}[ r_i(\tau), r_{i-1}(\tau), \phi_i, \theta_i]
\label{eq:02}
\end{eqnarray}

and the free energy of the system is: $F=\, -\beta ^{-1} \ln Z_N$. 
Importantly, the integration measure $\oint {D}r_i$ over the bp Fourier coefficients has to normalize the kinetic action. This condition sets the free energy zero and holds for any $\mu$. Hence, the free energy is independent of $\mu$ as expected for a classical system.  Furthermore, the normalization condition  yields the cutoff on the radial fluctuations  as explicitly shown in ref.\cite{io11a}. 
While the cutoff on the maximum fluctuation amplitudes is the same for all bps, note that, for a given molecule configuration, one may have a $r_i$ value which significantly differs from a neighbor $r_j$. This is in general the case and this accounts for the fact that a base can flip out of the stack thus causing local helical unwinding.

The cutoffs over the bending and twist angle integrations  are set to $\phi_{M}=\,\pi /4$ and $\theta_{M}=\,\pi /4$, respectively.  
Precisely, each bending angle $\phi_i$ between adjacent bps (see Fig.~\ref{fig:1}) is computed with respect to the average value for the preceding angle $< \phi _{i-1} >$ along the stack, i.e. $\phi_i =\, < \phi _{i-1} > + \phi_i^{fl}$ and the integration in Eq.~(\ref{eq:02}) is performed over the bending fluctuations $\phi_i^{fl}$ taken in the range $[- \phi_{M}, \phi_{M}]$.
This allows for the formation of kinks which locally bend the molecule axis, reduce the bending energy and increase the molecule flexibility.

Likewise, we use a recursive procedure which defines the twist angle $\theta_i$ of the $i-th$ bp with respect to the average $<\theta_{i - 1}>$ computed for the preceding bp along the axis. For the twist variable, one has to consider the increment $2\pi / h$ associated to the molecule helical conformation with $h$,  the number of bps per helix turn, being a tunable parameter of our computational method.
Accordingly, the twist angle is written as, \, $\theta_i =\, <\theta_{i - 1}>  + 2\pi / h + \theta_{i}^{fl}$ \, and the twist fluctuation variable $\theta_{i}^{fl}$  in Eq.~(\ref{eq:02}) is integrated in the range $[- \theta_{M}, \theta_{M}]$. 
The latter integration has to be performed for any value of the helical repeat $h$ taken in a physically meaningful range, $h \in \, [h_{min}, \, h_{max}]$ . In fact, given the absence of measurements for short chains, our method admits that short DNA  may have an helical repeat which differs from the experimental value, $h^{exp}=\, 10.4$,  usually considered for kilo-base B-DNA under physiological condition \cite{wang79}.  Accordingly we have taken a set of $n$ input values for $h$ centered around $h^{exp}$, i.e., $n=\, (h_{max} - h_{min})/\Delta h \,$ with $\Delta h$ being the partition mesh. Further, by integrating over the ensemble defined by Eqs.~(\ref{eq:02}),  we have derived a set of $n$ average twist conformations expressed by:

\begin{eqnarray}
< h >_{j}=\,\frac{2\pi N}{< \theta_N >} \, , \, \hskip 1cm  (j=\,1,...,n) ,
\label{eq:04}
\end{eqnarray}

where $< \theta_N >$ is the average twist for the last bp in the chain. 
The $< h >_{j}$'s denote the possible twist conformations for the short chain and, by minimizing the free energy over the computed $< h >_{j}$'s, the program derives the equilibrium average helical repeat ($< h >_{j*}$). This can be done under specific conditions defined e.g. by the presence of external forces, temperature and salt concentration in solution  \cite{cher11}.
It is understood that the $< h >_{j}$'s also contain the effects of the radial and bending fluctuations according to the integration recipe given in Eq.~(\ref{eq:02}). Certainly the accuracy of our calculation grows with the density of $n$-values taken in the $h$-range around $h^{exp}$. The following calculations are carried out in the window $h_{max} - h_{min}=\,4$ with a partition step $\, \Delta h=\,0.015625$.

\section{ Results}

First we apply the method to a set of homogeneous short sequences to highlight the effects of the chain length on the equilibrium properties. The potential parameters are taken in accordance with our previous works and are consistent with available information regarding thermodynamic and elastic DNA properties  i.e.,  $D_i=\,60 meV$, $b_i= 5 \AA^{-1}$,  $f_s=\,0.1$, $l_s=\,0.5 \AA$,  $K_S=\,10 mev \AA^{-2}$, $\rho_{i} \equiv \, \rho_{i, i-1} =\,1$, $\alpha_{i}\equiv \, \alpha_{i, i-1} =\, 2 \AA^{-1}$ . 
In particular these values permit to reproduce the range of experimental free energies, obtained by averaging the contributions of all dimers, in duplex DNA \cite{santa,io16b}. 

\begin{figure}
\includegraphics[height=7.0cm,width=7.0cm,angle=-90]{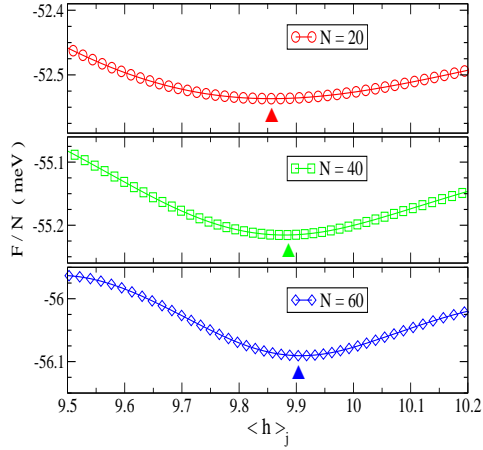}
\caption{\label{fig:2}(Color online)  
Room temperature free energy per base pair against the average helical repeat, defined in Eq.~(\ref{eq:04}), for three homogeneous molecules with (a) {} 20, \, (b) {} 40 and (c) {} 60 base pairs. The triangles indicate the equilibrium twist conformations corresponding to the free energy minima.
}
\end{figure}

The trend of the results displayed for the homogeneous chains would not change by assuming different sets of parameter values also used in DNA models   \cite{weber09,singh11,albu14}. 
Fig.~\ref{fig:2} shows the free energy per bp as a function of the $< h >_{j}$'s  for three chain lengths. The triangles mark the free energy minima yielding the equilibrium $< h >_{j*}$'s for the respective chains. A significant shift towards higher $< h >_{j*}$'s is found as one proceeds to consider longer chains. However, the $< h >_{j*}$'s increment as a function of $N$ is not linear suggesting that the equilibrium helical repeat should saturate at $\sim 10$, for chains in the range of hundreds bps (albeit not studied here). While the $< h >_{j*}$'s are the most probable twist conformations selected by free energy minimization, our plots indicate that, because of thermal fluctuations, the chains may assume also $< h >_{j}$'s conformations which are close to  $< h >_{j*}$ on the energy scale \cite{io15}.

\subsection{ Helical parameters}

The statistical averages in Eq.~(\ref{eq:02}) are performed to calculate the average distance between neighbor bps along the stack, \, $< d > = \, \frac{1}{N-1} \sum _{i=2}^{N} < \overline{d_{i,i-1}} > $ and the average bp fluctuation, \, $< R >= \, \frac{1}{N} \sum _{i=1}^{N} < r_i > $, with respect to the bare helix diameter.

\begin{figure}
\includegraphics[height=7.0cm,width=7.0cm,angle=-90]{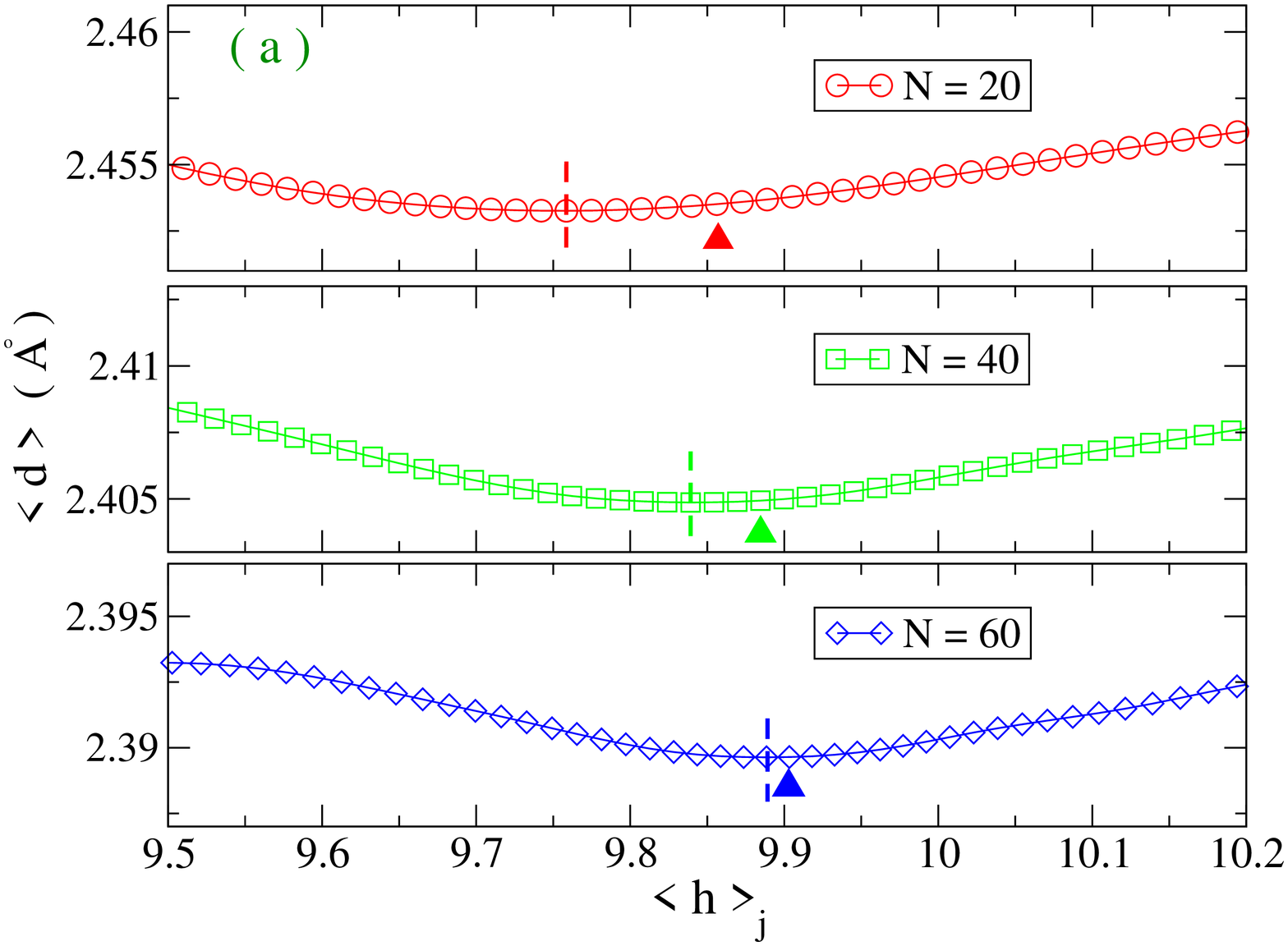}
\includegraphics[height=7.0cm,width=7.0cm,angle=-90]{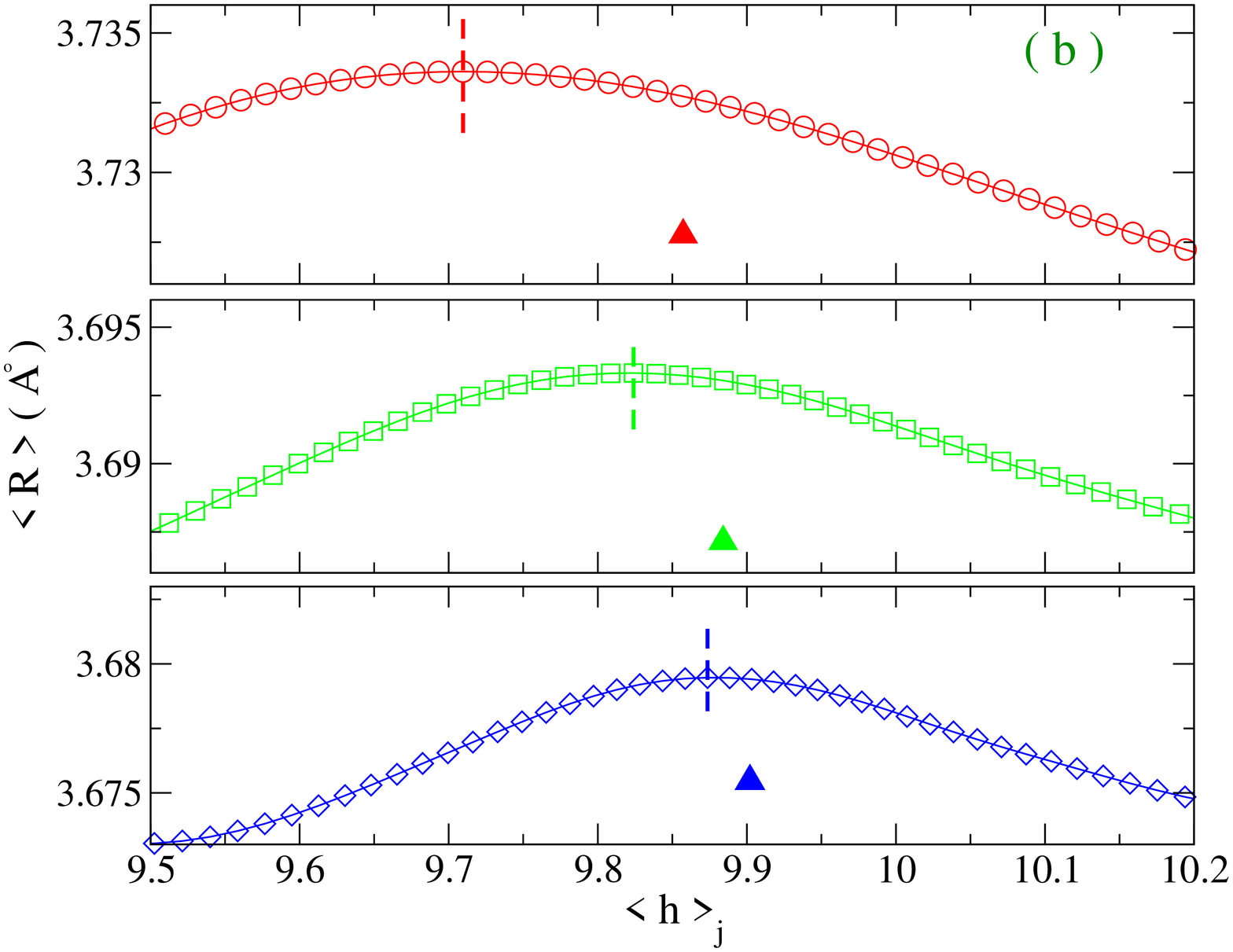}
\caption{\label{fig:3}(Color online)  
(a) {} Average helix elongation per dimer and (b) {} average base pair fluctuation respect to the bare diameter $R_0$, defined in the text, against the average helical repeat.  The same homogeneous sequences of Fig.~\ref{fig:2} are considered. The triangles indicate the equilibrium twist conformations. The dashed lines mark (a) the minimum base pair distances and (b) the maximum base pair fluctuations as a function of $< h >_{j}$.
}
\end{figure}

Both quantities may depend on the  twist conformation associated to a specific $< h >_{j}$ value. Then, we calculate $< d >$ and $< R >$  for the equilibrium twist $\,< h >_{j*}\,$  and further monitor their changes both in the untwisted ($< h >_{j}$  larger than  $< h >_{j*}$) and in the over-twisted ($ < h >_{j}$ smaller than  $< h >_{j*}$) conformations. The results are shown in Fig.~\ref{fig:3}.
First, we notice that there is an appreciable dependence of the helical parameters on the twist conformation, with $< d >$ and $< R >$ being non-monotonous versus $< h >_{j}$.
For very short chains, the minimum elongation and the maximum helix diameter (dashed lines) may not coincide with the equilibrium twist conformations (triangles) determined in Fig.~\ref{fig:2}, an effect attributed to the large fluctuational effects generally more pronounced in shorter chains. However, as the molecule length grows ($N=\,60$), the gap between the location of dashed line and triangle narrows thus signaling that the $< h >_{j*}$ conformation emerges as the one in which the average helix elongation is minimum and the average diameter is maximum. This condition should correspond to the most stable helix conformation driven by a balance between strong bonds along the molecule stack and breathing fluctuations which confer flexibility to the whole chain.

\subsection{  Persistence Length}

The orientational correlation between distant portions of the helical molecule may be lost due to disordering thermal fluctuations or to interactions with the surrounding solvent. Such effects vary both with the sequence specificities and type of solvent \cite{maiti18} thus concurring to determine the flexibility properties as measured by the $l_p$ \cite{save12}.  For semi-flexible polymers the latter is defined as the characteristic decay length of the correlation between two bond vectors, $\,\overline{d_{i}}\equiv \overline{d_{i,i-1}}$ and $\overline{d_{k}}\equiv \overline{d_{k,k-1}} \,$, that is:

$< \overline{d_{i}} \cdot \overline{d_{k}} > =\,\exp \bigl[-\bigl|\overline{d_{k}} - \overline{d_{i}}\bigr|/ l_p \bigr]$ \,.
 
For long molecules, the finite size effects are negligible as, for most $i-$ sites, the average correlation between $i$ and $k$ sites tends to vanish before approaching the chain end. Hence, the flexibility of most of the chain is well described by the exponentially decaying correlation function. The same concept of exponential decay (for the directional cosine of the bending angle between distant vectors) defines $l_p$ in the continuous WLC model \cite{gole12} which assumes an infinitely large number of bond lengths while the contour length is kept fixed. Although the WLC model is designed for long polymers, whose $l_p$ is expected to be lower than the contour length, the WLC relations have been applied to extract the apparent $l_p$ also in short chains, from analysis of the end-to-end distance and measurements of the DNA size, e.g. via the radius of gyration or the end-to-end distance \cite{archer08}. 

Thus, molecular dynamics computations of the bond vectors correlation and Monte Carlo calculations of the WLC mean square end-to-end distance \cite{tan15} deliver $l_p$ values smoothly increasing from $290$\AA {} to  $450$\AA {} for chain lengths growing in the range $N\in \,[10,\,50]$  whereas FRET data for a set of sequences with $N\in \,[15,\,21]$ yield $l_p=\,110$\AA {} through the analysis of the end-to-end distribution function \cite{archer08}. Further, for a chain with $50$ beads treated by the WLC model with stretching flexibility \cite{tan17}, the value $l_p=\,20$\AA {} has been fitted to the end-to-end distribution function of a coiled configuration, whose most probable radial distance $R^{M}_{e-e}$ is less than one half of its mean contour length $\overline L$, i.e. $R^{M}_{e-e} / \overline L \sim 0.45\,$. 

While all these cited $l_p$'s  are generally smaller than the standard value ($500$\AA) {} of kilo-base long DNA, the broad scattering of available estimates may suggest that the WLC picture itself breaks down when it comes to analyze the flexibility of short DNA chains.

For these reasons, we adopt here a microscopic definition of the apparent $l_p$ as an average over all the local site dependent $l_p(i)$ which, in turn, are obtained by averaging the bond vectors correlations over the ensemble in Eq.~(\ref{eq:02}). Then, in our method, $l_p$ reads:

\begin{eqnarray}
& &l_p(i) =\,  \frac{1}{< d >} \sum_{k=0}^{N-1-i} < \overline{d_{i,i-1}} \cdot \overline{d_{k,k-1}}> \, \nonumber
\\
& &l_p =\,  \frac{1}{N-1} \sum_{i=1}^{N-1}  l_{p}(i)  \, .
\label{eq:08}
\end{eqnarray}

\begin{figure}
\includegraphics[height=7.0cm,width=8.0cm,angle=-90]{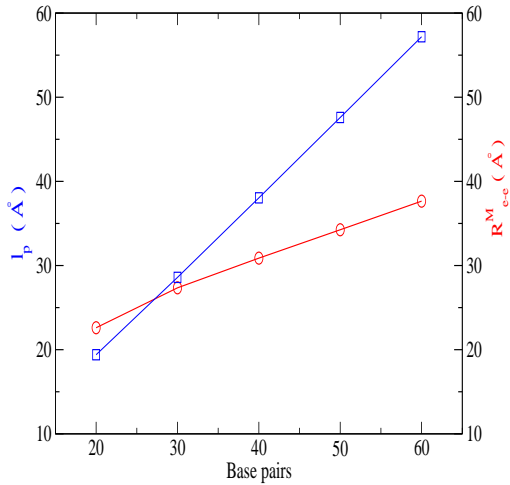}
\caption{\label{fig:4}(Color online)  
Apparent persistence length (left $y$-axis) and most probable end-to-end distance (right $y$-axis) for a set of homogeneous chains. The calculations are performed for the equilibrium twist conformation. The potential parameters are those used in Fig.~\ref{fig:2}.
}
\end{figure}

Computing Eq.~(\ref{eq:08}), we obtain the results shown in Fig.~\ref{fig:4} for a set of homogeneous fragments with $N\in \,[20,\,60]$ and with the same potential parameters as in Fig.~\ref{fig:2}. For each chain, the twist conformation is that selected by free energy minimization. It is found that: i) $l_p$ grows linearly with the chain length, ii) the obtained values are rather small pointing to a sizeable chain flexibility driven by entropic effects. 
This conclusion is corroborated by the most probable end-to-end distances ($R^{M}_{e-e}$) calculated for the same set and displayed in Fig.~\ref{fig:4} (right $y$-axis) \cite{io18a}. In fact $R^{M}_{e-e}$ markedly deviates from linearity as the chain length grows. Further, for each chain, $R^{M}_{e-e}$ is lower than one half of the mean contour length ($\overline L=\,(N-1) \cdot < d >\,$) indicating that coiled configurations have a large statistical weight also at short length scales.

While our estimates of the $l_p$'s in short chains appear consistent with some values reported by other studies, e.g. \cite{tan17}, it should be noticed that the latter works are based on a WLC model (without twist) in which $l_p$ appears as an adjustable parameter whereas, in our method, $l_p$ is directly obtained via computation of the correlation function built in the framework of a realistic Hamiltonian model for the twisted molecule.

\subsection{  Heterogeneity}

Clearly the $l_p$'s may depend on the chosen potential parameters and, in general, such values varies with the sequence specificities. To highlight this effect we consider the $30$ bps heterogeneous sequence with $15 \,AT$ and $15 \, GC$ bps:

\begin{eqnarray}
\,\, GGG AAA GGGGG AAAAAAA GG AA G AA G A GGG \, 
\label{eq:080}
\end{eqnarray}

Only the purine bases along one strand are given in Eq.~(\ref{eq:080}) as our model neglects differences in the stacking contributions arising e.g., from the dimers AA/TT,  AT/TA and TA/AT with the slash separating strands having opposite orientation  (analogously for the GC bps). 

On the other hand, the specific contributions due to the dimers AA/TT, AG/TC and GG/CC are distinguished in the stacking potential through the non-linear parameters 
$\rho_{i, i-1}$ and $\alpha_{i, i-1}$ (with the commas separating adjacent bps along the stack) which control amplitude and range of the bp fluctuation. 

In fact,  looking at the $V_2$ potential defined in Eq.~(\ref{eq:01}), as long as the condition $\,|r_i| - R_0 \ll (\alpha_{i, i-1})^{-1} \,$ holds for all bps, the molecule is stable and the effective stacking coupling is $\,K_S (1 + \rho_{i, i-1})$. Whenever, because of a fluctuation, the hydrogen bond of a specific bp is disrupted then the base moves out of the stack, the local coupling drops to $K_S$ and also the adjacent base tends to loose its bond with the complementary mate thus extending cooperatively the fluctuational bubble along the helix. This event yields an energetic gain which is proportional to $\rho_{i, i-1}$. Accordingly we attribute a larger anharmonic weight to AA/TT dimers (which are more prone to bending deformations than GG/CC dimers) taking larger values for the $\rho_{A, A}$ parameters. Conversely we assume $\,\alpha_{A, A} <  \alpha_{G, G} \,$ as the inverse lengths measure the amplitude of the bp fluctuation required to soften the stacking energy coupling. 

\begin{figure}
\includegraphics[height=7.0cm,width=8.0cm,angle=-90]{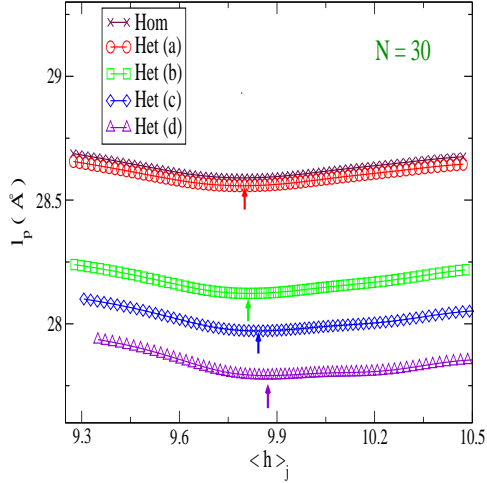}
\caption{\label{fig:5}(Color online)  
Persistence length for a fragment with 30 base pairs as a function of the twist conformation. Both the homogeneous case  and the heterogeneous sequence in Eq.~(\ref{eq:080}) are considered.  For the latter, four cases are assumed.  Het (a): Heterogeneity is introduced only through the Morse parameters ($D_{GC}=\,60 meV$, $b_{GC}= 5 \AA^{-1}\,$, $D_{AT}=\,40 meV$, $b_{AT}= 3 \AA^{-1}$ ) with all other potential parameters being equal to the homogeneous case. Het (b),(c),(d): Heterogeneity is introduced also via the $\rho_{i, i-1}$ and $\alpha_{i, i-1}\,$ parameters in Table~\ref{table:1}. The Morse parameters are kept as in Het (a). The arrows mark the equilibrium twist conformations.
}
\end{figure}

With these premises, we calculate by Eq.~(\ref{eq:08}) the $l_p$ for the sequence in Eq.~(\ref{eq:080}) against the twist conformation as displayed in  Fig.~\ref{fig:5}. The homogeneous case (potential parameters are those of the previous Figures) is plotted for comparison while the curve labeled by Het (a) is obtained by varying (with respect to the homogeneous chain) only the Morse parameters for the AT bps. This causes a minor effect on $l_p$,  however ascribable to the fact that the  averages in Eq.~(\ref{eq:08}) are carried out over the ensemble in Eq.~(\ref{eq:02}), weighed by the Boltzmann factor which also contains the Morse potential.

Instead, the bending flexibility is controlled by the stacking potential depending on the angular degrees of freedom. Once the chain heterogeneity is weighed in the model via the parameters $\rho_{i, i-1} \,$ and $\alpha_{i, i-1}\,$, we obtain a more pronounced effect on 
$l_p$ as shown by the curves Het (b),(c),(d) computed respectively for the input values in Table~\ref{table:1}. Such values are varied arbitrarily as there are no available data on short sequences to constrain the anharmonic parameters. Interestingly however,  the equilibrium helical repeat progressively shifts upwards (as marked by the arrows in the plot) by increasing the weight of the non-linear parameters while $l_p$ decreases. Thus the helical untwisting should be meant as an indicator of an increased molecule flexibility at the short length scales considered in this work.

\begin{table}
\begin{center}
 \begin{tabular}{|c| c| c| c| c| c| c|} 
 \hline
 & $\rho_{A,A}$ & $\alpha_{A,A}$ & $\rho_{A,G}$ & $\alpha_{A,G}$ & $\rho_{G,G}$ & $\alpha_{G,G}$ \\ [0.5ex] 
 \hline\hline
 (b) & 3 & 1 & 2 & 2 & 1 & 3 \\ 
 \hline
 (c) & 5 & 1 & 4 & 2 & 3 & 3 \\
 \hline
 (d) & 10 & 1 & 8 & 2 & 6 & 3 \\  [1ex] 
 \hline
\end{tabular}
\end{center}
\caption{Sets of non linear potential parameters used in Fig.~\ref{fig:5} to compute the $l_p$'s labeled by Het (b), (c), (d), respectively. $\rho_{i, i-1}$'s are dimensionless. $\alpha_{i, i-1}$'s are in units $\AA^{-1}$. }
\label{table:1}
\end{table}

\section{Conclusions}

I have focused on the persistence length of short fragments proposing a computational method which essentially treats the bp fluctuations as temperature dependent paths. 
Large amplitude fluctuations and pair breaking effects are included in the calculation which accounts for the formation of bubbles and flexible hinges along the chain. Furthermore, the Hamiltonian contains both the twist and the bending angle between adjacent bps along the stack hence, the path integration is carried out over all radial and angular fluctuations which concur to shape the helical molecule in solution. 
This feature is a significant advancement with respect to previous coarse-grained investigations which either neglect bp fluctuations or take the double stranded helix as a ladder.
The computational method yields thermodynamic quantities and helical parameters as a function of the specific twist conformation. 
Accordingly, I have calculated the correlation function which defines the microscopic $l_p$ for a set of short homogeneous fragments and obtained values which are markedly smaller than the standard $l_p$ for kilo-base long DNA. This suggests that, at short length scales, DNA maintains a remarkable flexibility also witnessed by its end-to-end distance which appears smaller than its average contour length. These results are in line with a body of experiments supporting the view that the intrinsic flexibility of short chains may be related to local bending of the bonds between adjacent bps. At the current stage however, there are not sufficient experimental information on the $l_p$ of short DNA to compare with the calculated values. I have also applied the method to a heterogeneous oligomer by tuning the potential parameters which weigh the anharmonic effects in the stacking potential. The persistence length, obtained as a function of the twist conformation, shows that an increased molecule flexibility is related to an appreciable untwisting of the helix.
I finally observe that the path integral technique is feasible for analysis of DNA properties in crowded environments as the stability of the helical configuration may depend on the specific confinement of the phase space available to the bps operated by the cell walls or by crowders which influence the free volume in the system.

\end{document}